\newif\ifarticle\articletrue
\ifarticle%
\documentclass[a4paper,DIV=11]{scrartcl}%
\pdfoutput=1
\usepackage[utf8]{inputenc}
\usepackage[USenglish]{babel}
\usepackage[T1]{fontenc}
\usepackage{amsthm}
\usepackage{fullpage}
\usepackage{authblk}
\usepackage{hyperref}
\date{}

\addtokomafont{title}{\normalfont\rmfamily}
\addtokomafont{disposition}{\rmfamily}
\newtheorem{theorem}{Theorem}

\newtheorem{definition}[theorem]{Definition}
\newtheorem{corollary}[theorem]{Corollary}
\newtheorem{lemma}[theorem]{Lemma}

\else%
\documentclass[a4paper,USenglish]{lipics}
\fi
\usepackage{epsfig}
\usepackage{graphics}
\usepackage{amsmath,setspace}
\usepackage{booktabs}
\usepackage{array}
\usepackage{multirow}
\usepackage{amssymb}
\usepackage{caption}
\usepackage{tikz}
\usetikzlibrary{backgrounds,arrows,shapes,snakes,decorations}

\begin{document}

\title{Parameterized Complexity of Critical Node Cuts}

\author[1]{Danny Hermelin}
\author[1]{Moshe Kaspi}
\author[2]{Christian Komusiewicz}
\author[1]{Barak Navon}
\affil[1]{Department of Industrial Engineering and Management, Ben-Gurion University, Israel\\
\href{mailto:hermelin@bgu.ac.il}{hermelin@bgu.ac.il}, \href{moshe@exchange.bgu.ac.il}{moshe@exchange.bgu.ac.il}, \href{baraknav@post.bgu.ac.il}{baraknav@post.bgu.ac.il}}
\affil[2]{Institut f\"{u}r Softwaretechnik und Theoretische Informatik, TU Berlin, Germany\\
\href{christian.komusiewicz@tu-berlin.de}{christian.komusiewicz@tu-berlin.de }
}
\ifarticle\else
\authorrunning{Hermelin et al.}
\fi 

\maketitle

\begin{abstract}

We consider the following graph cut problem called \textsc{Critical Node Cut (CNC)}: Given a graph $G$ on $n$ vertices, and two positive integers $k$ and $x$, determine whether $G$ has a set of $k$ vertices whose removal leaves $G$ with at most $x$ connected pairs of vertices. We analyze this problem in the framework of parameterized complexity. That is, we are interested in whether or not this problem is solvable in $f(\kappa) \cdot n^{O(1)}$ time  (\emph{i.e.}, whether or not it is \emph{fixed-parameter tractable}), for various natural parameters $\kappa$.  We consider four such parameters:
\begin{itemize}
\item The size $k$ of the required cut.
\item The upper bound $x$ on the number of remaining connected pairs.
\item The lower bound $y$ on the number of connected pairs to be removed.
\item The treewidth $w$ of $G$.
\end{itemize}
We determine whether or not \textsc{CNC} is fixed-parameter tractable for each of these parameters. We determine this also for all possible aggregations of these four parameters, apart from $w+k$. Moreover, we also determine whether or not \textsc{CNC} admits a polynomial kernel for all these parameterizations. That is, whether or not there is an algorithm that reduces each instance of \textsc{CNC} in polynomial time to an equivalent instance of size $\kappa^{O(1)}$, where $\kappa$ is the given parameter.

\end{abstract}

\section{Introduction}
\label{sec:introduction}

In $2013$ a polio virus struck Israel. The virus spread in alarming speed, creating a nationwide panic of parents concerned about the well-being of their children. It was obvious to the Israeli health department that vaccinating all Israeli children is not a practical solution in the given time frame. Thus it became clear that some areas of the country should be vaccinated first in order to stop the spread of the virus as quickly as possible. Let us represent a geographic area as a vertex of a graph, and the roads between areas as edges of the graph. In this setting, vaccinating an area corresponds to deleting a certain vertex from the graph. Thus, the objective of stopping the virus from spreading translates to minimizing the number of \emph{connected pairs} (two vertices which are in the same component) in the corresponding graph after applying the vaccination.

This scenario can be modeled by the following graph-theoretic problem which we call \textsc{Critical Node Cut} (\textsc{CNC}). In this problem, we are given an undirected simple graph $G$ and two integers $k$ and $x$. The objective is to determine whether there exists a set $C \subseteq V(G)$ of at most $k$ vertices in $G$, such that the graph $G-C$ which results from removing $C$ from $G$, contains at most $x$ connected pairs. In this sense, the cut $C$ is considered \emph{critical} since removing it from $G$ leaves few (at most $x$) connected pairs. For convenience purposes, throughout the paper we will count \emph{ordered} connected pairs; \emph{i.e.}, pairs $(u,v) \in V(G) \times V(G)$, $u \neq v$, where $u$ and $v$ belong to same connected component in $G-C$.

The goal of \textsc{CNC} is thus, roughly speaking, to destroy the connectivity of a given graph as much as possible given a certain budget for deleting vertices. From this point of view, \textsc{CNC} fits nicely to the broad family of 
\emph{graph-cut problems}. Graph-cut problems have been studied widely and are among the most fundamental problems in algorithmic research. 
Examples include \textsc{Min Cut}, \textsc{Max Cut}, \textsc{Multicut}, \textsc{Multiway Cut}, \textsc{Feedback Vertex Set}, and \textsc{Vertex Cover} (see \emph{e.g.}~\cite{gj79} for definitions of these problems). The latter is the special case of \textsc{CNC} with $x=0$. Since \textsc{Vertex Cover} is arguably the most important problem in the theory of algorithmic design for NP-hard problems, \textsc{CNC} provides a natural test bed to see which of the techniques from this theory can be extended, and to what extent.

\paragraph*{Previous Work and Applications.}

The \textsc{CNC} problem has been studied from various different angles. The problem was shown to be NP-complete in~\cite{3} (although its NP-completeness follows directly from the much earlier NP-completeness result for \textsc{Vertex Cover}). In trees, a weighted version of \textsc{CNC} is NP-complete whereas the unweighted version can be solved in polynomial time~\cite{6}. The case of bounded treewidth can be solved using dynamic programming in $O(n^{w+1})$ time, where $n$ is the number of vertices in the graph and $w$ is its treewidth~\cite{7}. Local search~\cite{3} and simulated annealing~\cite{10} were proposed as heuristic
algorithms for \textsc{CNC}. Finally, in~\cite{2} an approximation algorithm based on randomized rounding was developed.

Due to its generic nature, the \textsc{CNC} problem has been considered above in various different application settings. One example application is the virus vaccination problem discussed above~\cite{3}. Other interesting applications include protecting a computer/communication network from corrupted nodes, analyzing anti-terrorism networks~\cite{21}, measuring centrality in brain networks~\cite{22}, insulin signaling~\cite{37}, and protein-protein interaction network analysis~\cite{38}.

\paragraph*{Our Results.}
From reviewing the literature mentioned above, it is noticeable that an analysis of \textsc{CNC} from the perspective of parameterized complexity~\cite{DowneyFellows1999} is lacking. The purpose of this paper is to remedy this situation. We examine \textsc{CNC} with respect to four natural parameters along with all their possible combined aggregations. The four basic parameters we examine are:
\begin{itemize}
\item The size $k$ of the solution (\emph{i.e.} the critical node cut) $C$.
\item The bound $x$ on the number of connected pairs in the resulting graph $G-C$.
\item The number of connected pairs $y$ to be removed from $G$; if $G$ is connected with $n$ vertices then $y=n(n-1)-x$.
\item The treewidth $w$ of $G$.
\end{itemize}
Table~\ref{tab:results} summarizes all we know regarding the complexity of \textsc{CNC} with respect to these four parameters and their aggregation.
\begin{table}[t]
\small
\centering
\begin{tabular}{|>{\centering\arraybackslash}p{1cm}|>{\centering\arraybackslash}p{1cm}|
>{\centering\arraybackslash}p{1cm}|>{\centering\arraybackslash}p{1cm}|c|c|}
\hline
\multicolumn{4}{|c|}{Parameter} & \multicolumn{2}{|c|}{Result}\\
\hline
$k$ & $x$ & $y$ & $w$ & FPT & P-Kernel\\
\hline
\checkmark &  &  &  & \quad \textsc{NO} (Thm.~\ref{thm:k}) \quad & \quad \textsc{NO} (Thm.~\ref{thm:k}) \quad\\
\hline
& \checkmark &  &  & \textsc{NO} & \textsc{NO}\\
\hline
&  & \checkmark &  & \textsc{YES} (Thm.~\ref{thm:y}) & \textsc{NO} (Thm.~\ref{thm:y+k+w})\\
\hline
&  &  & \checkmark & \textsc{NO} (Thm.~\ref{thm:w}) & \textsc{NO} (Thm.~\ref{thm:y+k+w})\\
\hline
\checkmark & \checkmark &  &  & \textsc{YES} (Thm.~\ref{thm:k+x}) & \textsc{YES} (Thm.~\ref{thm:k+x_ker})\\
\hline
\checkmark &  & \checkmark &  & \textsc{YES} (Thm.~\ref{thm:y}) & \textsc{NO} (Thm.~\ref{thm:y+k+w})\\
\hline
\checkmark &  &  & \checkmark & \textbf{?} & \textsc{NO} (Thm.~\ref{thm:y+k+w})\\
\hline
& \checkmark & \checkmark &  & \textsc{YES}  & \textsc{YES}\\
\hline
& \checkmark &  & \checkmark & \textsc{YES} (Thm.~\ref{thm:w+x}) & \textsc{NO}\\
\hline
&  & \checkmark & \checkmark & \textsc{YES} (Thm.~\ref{thm:y}) & \textsc{NO} (Thm.~\ref{thm:y+k+w})\\
\hline
\checkmark & \checkmark & \checkmark &  & \textsc{YES}  & \textsc{YES} \\
\hline
\checkmark & \checkmark &  & \checkmark & \textsc{YES} (Thm.~\ref{thm:k+x}) & \textsc{YES} (Thm.~\ref{thm:k+x_ker})\\
\hline
\checkmark &  & \checkmark & \checkmark & \textsc{YES} (Thm.~\ref{thm:y}) & \textsc{NO} (Thm.~\ref{thm:y+k+w})\\
\hline
& \checkmark & \checkmark & \checkmark & \textsc{YES}  & \textsc{YES}\\
\hline
\checkmark & \checkmark & \checkmark & \checkmark & \textsc{YES}  & \textsc{YES} \\
\hline
\end{tabular}
\centering
\caption{Summary of the complexity results for \textsc{Critical Node Cut}.}
\label{tab:results}%
\end{table}
Let us briefly go through some of the trivial results given in the table above. First note that \textsc{CNC} with $x=0$ is precisely the \textsc{Vertex Cover} problem, which means that \textsc{CNC} is not in FPT (and therefore has no polynomial kernel) for parameter $x$ unless P=NP. This also implies that the problem is unlikely to admit a polynomial kernel even when parameterized by $w+x$, since such a kernel would imply a polynomial kernel for \textsc{Vertex Cover} parameterized by the treewidth~$w$ which is known to cause the collapse of the polynomial hierarchy~\cite{Bodlaender-et-al2009a,Drucker2012}. Next, notice that if our input graph $G$ has no isolated vertices, we have $x+y = \Omega(n)$, and therefore \textsc{CNC} is FPT and has a polynomial kernel (as isolated vertices can safely be discarded). This of course means that the same applies for parameters $k+x+y$, $x+y+w$, and $k+x+y+w$.

Our first result, stated in Theorem~\ref{thm:k}, shows that \textsc{CNC} parameterized by $k$ is W[1]-hard. Thus, CNC is unlikely to have an FPT algorithm under this parameterization. We then show in Theorem~\ref{thm:k+x} and Theorem~\ref{thm:k+x_ker}, that when considering $x+k$ as a parameter, we can extend two classical \textsc{Vertex Cover} techniques to the \textsc{CNC} problem. Our main technical result is stated in Theorem~\ref{thm:w}, where we prove that \textsc{CNC} is W[1]-hard with respect to $w$, the treewidth of the input graph. This is somewhat surprising since not many graph cut problems are known to be W[1]-hard when parameterized by treewidth. Also, the result complements nicely the $O(n^{w+1})$-time algorithm of~\cite{7} by showing that this algorithm cannot be improved substantially. 
We complement this algorithm from the other direction by showing in Theorem~\ref{thm:w+x} that \textsc{CNC} can be solved in $f(w+x)\cdot n^{O(1)}$ time. Finally, we show in Theorem~\ref{thm:y} and Theorem~\ref{thm:y+k+w} that \textsc{CNC} is FPT with respect to~$y$, and has no polynomial kernel even if $y$, $w$, and $k$ are taken together as parameters.

\paragraph*{Related Work.}

This paper belongs to a recent extensively explored line of research in parameterized complexity where various types of graph cut problems are analyzed according to various natural problem parameterizations. This line of research can perhaps be traced back to the seminal paper of Marx~\cite{Marx2006} who studied five such problems, and in the process introduced the fundamental notion of \emph{important separators}. This paper paved the way to several parameterized results for various graph cut problems, including \textsc{Multicut}~\cite{BousquetEtAl2011,Guillemot2011,KratschEtAl2012,Marx2006,MarxEtAl2013,MarxRazgon2011,Xiao2010}, \textsc{MultiwayCut}~\cite{CaoEtAl2013,ChenEtAl2009,CyganEtAl2013,Guillemot2011,Marx2006,Xiao2010}, and \textsc{Steiner Multicut}~\cite{BringmannHML15}. A particularly closely related problem to \textsc{CNC} is the so-called \textsc{Vertex Integrity} problem where we want to remove $k$ vertices from a graph such that the largest connected component in the remaining graph has a bounded number of vertices. Fellows and Stueckle~\cite{FellowsStueckle1989} were the first to analyze this problem from a parameterized point of view. We refer the reader to~\cite{DrangeDH14} for a detailed overview of the known results on this problem.  

\section{Parameters \boldmath{$k$} and \boldmath{$k+x$}}
\label{sec:k and k+x}

We now consider the parameters $k$ and $k+x$ for \textsc{CNC}. 
We will show that the problem is W$[1]$-hard for the former parameterization, while for the latter is in FPT and admits a polynomial kernel; the proofs are deferred to the appendix. 
\begin{theorem}
\label{thm:k}
\textsc{Critical Node Cut} is \textnormal{W[1]}-hard with respect to $k$.
\end{theorem}
\begin{proof}
We present a reduction\footnote{A somewhat similar reduction for related problems was given for example by~Marx~\cite{Marx2006}. } from the \textsc{Clique} problem: Given a graph $G$ on $n$ vertices, and a parameter $\ell$, determine whether $G$ has a pairwise adjacent subset of $\ell$ vertices. Let $(G,\ell)$ be an instance of \textsc{Clique}. We construct $H$, the graph of our \textsc{CNC} instance, as follows: We replace each edge in $G$ by a simple \emph{edge-gadget}. This is done by replacing the edge by $n$ parallel edges, and then subdividing each of the new edges once. The newly inserted subdivision vertices are referred to as \emph{dummy vertices}. We then add an edge in $H$ between each pair of nonadjacent vertices of $G$. Finally, we set $k:=\ell$.

We claim that the graph $G$ has a clique of size $\ell$ if and only if the graph $H$ has $k=\ell$ vertices whose removal deletes $$y=k(k-1) + 2k(N-k)+\binom{k}{2}n\cdot \left(\binom{k}{2}n-1\right) + 2\binom{k}{2}n\left(N-k-\binom{k}{2}n\right)$$ connected pairs in $H$, where $N:=|V(H)|$. 

We begin with the easier direction: Suppose $C$ is a clique of size
$\ell$ in $G$, and let~$D(C)$ denote the~$\binom{k}{2}n$ many dummy vertices that have both
neighbors in~$C$. Removing $C$ in $H$ results in the deletion of $y$
connected pairs from $H$: A total of
\begin{itemize}
\item $k(k-1)$ connected pairs which involve only vertices of $C$,
\item $2k(N-k)$ pairs which involve one vertex from~$C$ and one vertex
  from~$V(H)\setminus C$,
\item $\binom{k}{2}n\cdot \left(\binom{k}{2}n-1\right)$ which involve
  only vertices from~$D(C)$, and
\item $2\binom{k}{2}n\left(N-k-\binom{k}{2}n\right)$ connected pairs
  that involve one vertex of~$D(C)$ and one vertex of~$|V(H)|\setminus (C\cup D(C))$.
\end{itemize}

Conversely, suppose that $C$ is a cut that removes $y$ connected pairs
in $H$. Observe that if $C$ contains a subset $C'\subseteq C$ of dummy
vertices, then we can replace $C'$ with an arbitrary equally sized set
of non-dummy vertices without decreasing $y$. Thus, we can assume that
$C$ contains only non-dummy vertices. Furthermore, notice that when we
remove a non-dummy vertex~$v$ (\emph{i.e.}, a vertex of $G$), then the
only connected pairs that are deleted are the ones which either
involve~$v$ or possibly dummy vertices that are neighbors of~$v$. This
is because every pair of vertices from $G$ is either connected by an
edge or by an edge-gadget in $H$. Now, for \emph{every} set~$C$ of
size~$k$, removing~$k$ deletes exactly $k(k-1)+2k(N-k)$ connected
pairs that contain exactly one vertex from~$C$. Thus, the only way to
delete $y$ connected pairs in $H$ is to isolate~$\binom{k}{2}n$ dummy
vertices. This is only possible if we remove~$k$ vertices which are
pairwise connected by edge-gadgets; these correspond to $\ell=k$
vertices that form a clique in $G$. 
\end{proof}

We next show that the above result holds also for some restricted
subclasses. A \emph{split graph} is a graph in which the vertices can
be partitioned into a clique and an independent set. We slightly
modify the construction in the proof of Theorem~\ref{thm:k} by adding
all the edges missing between every pair of non-dummy vertices. In
this way, the vertices of $G$ form a clique and the dummy vertices
form an independent set, while all arguments in the proof above still
hold. For a fixed integer $d \geq 1$, a graph is called
\emph{$d$-degenerate} if each of its subgraphs has a vertex with a
degree of at most $d$. For $d=1$ (\emph{i.e.} a forest), the CNC
problem has a polynomial algorithm based on dynamic
programming~\cite{6}. We modify the construction in the proof above by
subdividing all the edges except those that are adjacent to dummy
vertices. This results in a $2$-degenerate graph, and also a bipartite
graph with one side containing all vertices of $G$ and the other
containing all the (old and ``new'') dummy vertices. Let~$N$ denote
the number of vertices in the new graph~$H'$. Again,~$H'$ has a set
of~$\ell$ vertices such that removing these vertices deletes~$y$ pairs
if and only if~$G$ has a clique of size~$\ell$. This can be seen as
follows. The only vertices that can be come disconnected by removing
at most~$\ell$ vertices are the dummy vertices. Thus, there is a
solution in which no dummy vertex is removed: If its neighbors are
both removed, then removing~$v$ does not destroy any other
pair. Otherwise, the same number of connected pairs can be deleted by
removing one of the neighbors instead, as they are still connected
even if~$v$ is removed. Thus, all removed vertices correspond again to
vertices of the \textsc{Clique} instance. Now the remaining part of
the proof is exactly the same as in Theorem~\ref{thm:k} except that
now removing a pair of vertices that are nonadjacent in the
\textsc{Clique} instance isolates one ``new'' vertex, but this is
still much less than the~$n$ isolated dummy vertices in the case of
vertices that are adjacent in the \textsc{Clique} instance.
\begin{corollary}\label{cor:split}
\textsc{Critical Node Cut} remains \textnormal{W[1]}-hard with respect to $k$ even if the input graph is split, bipartite, or $d$-degenerate for any fixed $d \geq 2$.
\end{corollary}

We next consider the parameter $k+x$. We will show that the basic techniques known for the case of $x=0$, \emph{i.e.}, the \textsc{Vertex Cover} problem, can be extended to the case where $x > 0$. First, a simple branching strategy can be developed into an FPT algorithm for the parameter $k+x$.
\begin{theorem}
\label{thm:k+x}%
\textsc{Critical Node Cut} is \textnormal{FPT} with respect to $k+x$.
\end{theorem}
\begin{proof}
Let $(G,k,x)$ be an instance of \textsc{CNC}, and let $n$ denote the number of vertices in $G$. Observe that if there exists a $C \subseteq V(G)$ of size $k$ such that $G-C$ has at most $x$ connected pairs, then $G-C$ has at most $x$ edges. Using this observation we will solve an auxiliary problem in order to determine whether $(G,k,x)$ has a solution. The objective of our auxiliary problem is to determine whether there exist $k$ vertices $C' \subseteq V(G)$ such that $G-C'$ has at most $x$ edges. Observe that we can solve this problem using the bounded search tree technique. For an arbitrary edge  $\,e=\{u,v\} \in E(G)$, we recursively branch on each of the following instances $(G-u,k-1,x)$, $(G-v,k-1,x)$, and~$(G-e,k,x-1)$. Here,~$G-e$ denotes the graph obtained by removing the edge~$e$ (and not the graph obtained by removing the two endpoints of~$e$). This process continues recursively until no edges remain, or~$k=x=0$.

An important attribute of this search tree algorithm is that it enumerates all the possible minimal solutions. Therefore, after applying the above algorithm, we obtain the set $\mathcal{C'}$ of all the minimal solutions to our auxiliary problem. If there exists a solution $C$ to our \textsc{CNC} instance, then $C$ is also a solution for the auxiliary problem but not necessarily a minimal solution. We apply brute force on each minimal solution in $\mathcal{C'}$ to check if it is possible to extend it into a solution for \textsc{CNC}. If this is not possible for every solution in $\mathcal{C'}$, then our \textsc{CNC} instance $(G,k,x)$ has no solution.

To analyze the running time of the algorithm described above, note that solving our auxiliary problem requires $3^{x+k} \cdot n$ time, and the size of the set of all minimal solutions $\mathcal{C'}$ generated by this algorithm is bounded by $3^{x+k}$. Let us next bound the running-time required for processing each minimal solution: Assume~$C' \in \mathcal{C'}$ contains $k_1$ vertices, leaving us a budget of $k_2=k-k_1$ vertices for our critical node cut. Now we can discard isolated vertices in $G-C'$ since these are irrelevant, and obtain a graph with at most $2x$ vertices. If~$k_2>2x$, then all vertices can be deleted. Otherwise, checking each possible way to extend $C'$ into a critical node cut requires $O(\binom{2x}{k_2}x^2+n)$ time, and the running time of the entire algorithm is $O(3^{x+k}(x^{k+2}+n))$. 
\end{proof}

The running time can be improved by using a more
elaborate approach in the last step. For example, isolated edges can
be dealt with in a dynamic programming subroutine. Then the remaining
instance on which brute-force has to be applied has at most~$1.5x$
vertices. Next, we show that a simple ``high-degree rule'' leads to a
polynomial kernel.
\begin{theorem}\label{thm:k+x_ker}%
\textsc{Critical Node Cut} has a polynomial kernel with respect to $k+x$.
\end{theorem}
\begin{proof}
Let $(G,k,x)$ be an instance of \textsc{CNC}. We will show a polynomial reduction from $(G,k,x)$ to an equivalent instance $(G',k',x)$ of \textsc{CNC} such that the number of vertices in $G'$ is polynomial in $k+x$. Our algorithm is in the same spirit of Buss's classical \textsc{Vertex cover} kernel~\cite{28}. We construct $G'$ by iteratively applying a high-degree rule until it can no longer be applied: Start with $k'=k$. Note that a vertex with more than $k'+\sqrt{x}$ neighbors must be in any critical node cut of size $k$. Thus, our high-degree rule checks if there is a vertex with degree at least $k'+\sqrt{x}+1$ in the graph, and if so, it removes it and decreases $k'$ by one. Once all high degree vertices are removed, we remove all isolated vertices to obtain $G'$. Clearly, this reduction runs in polynomial time. Moreover, $(G,k,x)$ is a yes-instance iff $(G',k',x)$.

Let us next bound the number of vertices in $G'$. Suppose there is a $C \subseteq V(G')$ of at most $k'\leq k$ vertices where $G'-C$ has at most $x$ connected pairs. We partition the remaining vertices of $G'$ into two sets $A$ and $B$, $A \cup B = V(G) \setminus C$. The set $A$ contains all isolated vertices in $G'-C$, while $B$ contains the non-isolated vertices of $G'-C$. Clearly $|B| \leq x$, since otherwise there would be more than $x$ connected pairs in $G'-C$. Now, since $G'$ has no isolated vertices by construction, each vertex of~$A$ is adjacent to at least one vertex of $C$. Moreover, since each vertex of~$C$ has at most $k+\sqrt{x}$ neighbors, and $C$ has at most $k$ vertices, this implies that $|A| \leq k(k+\sqrt{x})$. Thus,
$$
|V(G')| = |A|+|B|+|C| \leq k(k+\sqrt{x})+x+k 
$$
and the theorem is proved. 
\end{proof}

\section{Parameter \boldmath{$w$}}
\label{sec:w}

\newcommand{\X}{n^4}         
\newcommand{\Y}{n^9}       
\newcommand{\Z}{n^{16}}       
\newcommand{\XX}{n^{8}}      
\newcommand{\YY}{n^{18}}      
\newcommand{\ZZ}{n^{32}}      
\newcommand{\A}{\ell^2}    
\newcommand{\B}{\ell^4}    
\newcommand{\C}{\ell^7}    
\newcommand{\LA}{\ell^3}   

In this section we will show that \textsc{CNC} is unlikely to be fixed-parameter tractable when parameterized by~$w$. This implies that we cannot substantially improve on the $O(n^{w+1})$ algorithm of~\cite{7}. Since we will not directly use the notion of treewidth and tree decompositions, we refer to~\cite{36} for their definition. 
\begin{theorem}
\label{thm:w}%
\textsc{Critical Node Cut} is \textnormal{W[1]}-hard with respect to the treewidth~$w$ of the input graph.
\end{theorem}

Our proof of the theorem above is via the well-known multicolored clique technique~\cite{Fellows-et-al2009} which utilizes generic gadget structure to construct a reduction from the W[1]-complete \textsc{Multicolored Clique} problem: Given an undirected simple graph $G$ with $n$ vertices and $m$ edges, a coloring function $c:V(G) \to \{1,\ldots,\ell\}$ of the vertices of $G$, and a parameter $\ell$, determine whether $G$ has a clique which includes exactly one vertex from each color. Throughout the section we use $(G,c,\ell)$ to denote an arbitrary input to \textsc{Multicolored Clique}. As usual in parameterized reductions, we can assume that $n$ and $\ell$ are sufficiently larger than any fixed constant, and $n$ is sufficiently larger than $\ell$.

In the multicolored clique technique, we construct \emph{selection} gadgets which encode the selection of vertices and edges of $G$ (one per each color class and pair of color classes, respectively), and \emph{validation} gadgets which ensure that the vertices and edges selected indeed form a clique in $G$. In our reduction below, we will force any feasible solution to delete a large number of vertices from the constructed \textsc{CNC} instance in order to reach the required bound on the number of remaining connected pairs. We will ensure that such a solution always leaves $4\binom{\ell}{2}$ very large components which encode the selection of $\binom{\ell}{2}$ edges in $G$. The bound on the number of connected pairs will require all these huge components to have equal size, which in turn can only happen if the edges selected in $G$ are edges between the same set of $\ell$ vertices (implying that these $\ell$ vertices form a clique in~$G$). In what follows, we use $(H,k,x)$ to denote the instance of \textsc{CNC} that we construct, where $H$ is the input graph, $k$ is the size of the required cut, and $x$ is the bound on the number of connected pairs. Note that for our proof to go through, we will also need to show that the treewidth of $H$ is bounded by some function in $\ell$. \\

\noindent \textbf{Connector gadgets:} To each vertex $u \in V(G)$, we assign two unique integer identifiers: $low(u) \in \{1,\ldots,n\}$ and $high(u) \in \{n+1,\ldots,2n\}$, where $high(u)= 2n+1-low(u)$. Our selection gadgets are composed from gadgets which we call \emph{connector gadgets}. A connector gadget \emph{corresponds} to a vertex of $G$, and can be of \emph{low~order} or \emph{high~order}. A low~order connector gadget corresponding to a vertex $u \in V(G)$ consists of a clique of size $\B$ and an independent set of size $\Z+low(u)$ which have all edges between them; that is, it is a complete split graph on these two sets of vertices. Similarly, a high~order connector gadget corresponding to $u\in V(G)$ is a complete split graph on a clique of size $\B$ and an independent set of size $\Z+high(u)$.

We refer to the clique in a connector gadget as the \emph{core} of the gadget, and to the remaining vertices as the \emph{guard} of the gadget. Only vertices in the core will be adjacent to vertices outside the gadget. Notice that the huge independent set in the core contributes to a large number of connected pairs in $H$, and one can delete all these connected pairs only by adding all core vertices to the solution cut. Below we use this property to help us control solutions for our \text{CNC} instance. \\

\begin{figure}[t]
\begin{center}

  \begin{tikzpicture}[vert/.style={circle,thick,draw,inner sep=0.7pt,minimum size=1.5mm},
    svert/.style={circle,thick,fill=black,draw,inner sep=0pt,minimum size=1.5mm},x=0.8cm,y=0.95cm]
    \tikzstyle{mybox} = [draw, very thick,
    rectangle, rounded corners, minimum size=10.5mm, inner ysep=1pt]

    \tikzstyle{myell} = [draw, ellipse, very thick, inner xsep=4pt, inner ysep=4pt]

    \tikzstyle{myellv} = [draw, ellipse, very thick, inner xsep=5pt, inner ysep=15pt]

    \tikzstyle{myellh} = [draw, ellipse, very thick, inner ysep=5pt, inner xsep=20pt]


    \pgfmathsetmacro{\bracexl}{1.29}
    \pgfmathsetmacro{\bracexr}{2.71}
    \pgfmathsetmacro{\bracey}{1.38}

    \pgfmathsetmacro{\bracexdiff}{3}
    \pgfmathsetmacro{\braceydiff}{0.73}

    \draw [decorate, thick, decoration={brace}] (\bracexr,\bracey) -- (\bracexl,\bracey);
    \draw [decorate, thick, decoration={brace}] (\bracexl,\bracey-\braceydiff) -- (\bracexr,\bracey-\braceydiff);

    \draw [decorate, thick, decoration={brace}] (\bracexr+\bracexdiff+0.1,\bracey) -- (\bracexl+\bracexdiff-0.1,\bracey);
    \draw [decorate, thick, decoration={brace}] (\bracexl+\bracexdiff-0.1,\bracey-\braceydiff) -- (\bracexr+\bracexdiff+0.1,\bracey-\braceydiff);

    \draw [decorate, thick, decoration={brace}] (\bracexr+\bracexdiff+\bracexdiff+0.22,\bracey) -- (\bracexl+\bracexdiff+\bracexdiff-0.22,\bracey);
    \draw [decorate, thick, decoration={brace}] (\bracexl+\bracexdiff+\bracexdiff-0.22,\bracey-\braceydiff) -- (\bracexr+\bracexdiff+\bracexdiff+0.22,\bracey-\braceydiff);

    \draw [decorate, thick, decoration={brace}] (0.55,0.35) -- (-0.55,0.35);
    \node  (unum) at (0,0) {$\ell^2$};

    \draw [decorate, thick, decoration={brace}] (-2,0.3) -- (-2,1.7);
    \node  (vnum) at (-2.5,1) {$n^9$};
    
    \draw [decorate, thick, decoration={brace}] (10.7,-1.3) -- (9.3,-1.3);
    \node  (unum) at (10,-1.55) {$n^4$};

        \node  (l) at (2,1) {
      $\ell^4$
    };

    \node  (4l) at (5,1) {
      $\ell^7$
    };

    \node  (ll) at (8,1) {
      $\ell^4$
    };





    \pgfmathsetmacro{\bracexlg}{1}
    \pgfmathsetmacro{\bracexrg}{3}
    \pgfmathsetmacro{\braceyg}{4.6}

    \pgfmathsetmacro{\bracexdiffg}{3}
    \pgfmathsetmacro{\braceydiffg}{0.86}


    \draw [decorate, thick, decoration={brace}] (\bracexlg,\braceyg) -- (\bracexrg,\braceyg);
    \draw [decorate, thick, decoration={brace}] (\bracexlg+6,\braceyg) -- (\bracexrg+6,\braceyg);

    \node  (n5h) at (2,5) {
      $n^{16}+high(u_{1})$
    };

    \node  (n5l) at (8,5) {
      $n^{16}+low(u_{1})$
    };

    \pgfmathsetmacro{\braceyg}{-2.6}
    \draw [decorate, thick, decoration={brace}] (\bracexrg,\braceyg) -- (\bracexlg,\braceyg);
    \draw [decorate, thick, decoration={brace}] (\bracexrg+6,\braceyg) -- (\bracexlg+6,\braceyg);

    \node  (n5lb) at (2,-3) {
      $n^{16}+low(u_{1})$
    };

    \node  (n5hb) at (8,-3) {
      $n^{16}+high(u_{1})$
    };

    \node [mybox] (Ah) at (2,2) {
     \small $A^j_{high}[u_{1}]$
    };

    \node [myell] (Bh) at (2,4) {
     \small $B^j_{high}[u_{1}]$
    };

    \node [mybox] (u) at (0,1) {
     \small $U$
    };

    \node [mybox] (Al) at (2,0) {
     \small $A^j_{low}[u_{1}]$
    };

    \node [myell] (Bl) at (2,-2) {
     \small $B^j_{low}[u_{1}]$
    };

    \node [mybox] (Vh) at (5,2) {
     \small $V_{high}[i,j]$
    };

    \node [mybox] (Vl) at (5,0) {
     \small $V_{low}[i,j]$
    };

    \node [mybox] (Auvl) at (8,2) {
     \small $A^{u_{1}}_{low}[u_{1},u_{2}]$
    };

    \node [myell] (Buvl) at (8,4) {
     \small $B^{u_{1}}_{low}[u_{1},u_{2}]$
    };

    \node [mybox] (Auvh) at (8,0) {
     \small $A^{u_{1}}_{high}[u_{1},u_{2}]$
    };

    \node [myell] (Buvh) at (8,-2) {
     \small $B^{u_{1}}_{high}[u_{1},u_{2}]$
    };
    \node [mybox] (Avul) at (12,2) {
     \small $A^{u_{2}}_{low}[u_{1},u_{2}]$
    };
    \node [myell] (Bvul) at (12,4) {
     \small $B^{u_{2}}_{low}[u_{1},u_{2}]$
    };
    \node [mybox] (Avuh) at (12,0) {
     \small $A^{u_{2}}_{high}[u_{1},u_{2}]$
    };

    \node [myell] (Bvuh) at (12,-2) {
     \small $B^{u_{2}}_{high}[u_{1},u_{2}]$
    };


    \node [myellv] (udummy) at (-1.5,1) {};

    \node [svert,label=above:{$\{ u_{1},u_{2} \}$}] (uv) at (10,1) {};

    \node [myellh] (uvdummy) at (10,-1) {};

    \draw [dotted,thick] (u) -- (udummy);
    \draw [dotted,thick] (uv) -- (uvdummy);
    \draw [dotted,thick] (u) -- (Ah)--(Vh)--(Auvl)--(uv)--(Avul)--(Bvul);
    \draw [dotted,thick] (u) -- (Al)--(Vl)--(Auvh)--(uv)--(Avuh)--(Bvuh);

    \draw [dotted,thick] (Ah)--(Bh);
    \draw [dotted,thick] (Al)--(Bl);

    \draw [dotted,thick] (Auvh)--(Buvh);
    \draw [dotted,thick] (Auvl)--(Buvl);

\end{tikzpicture}

\caption{The connection of selection gadgets via a validation gadget. In the example, we consider a vertex $u_{1} \in V(H)$ with $c(u_{1})=i$ which is adjacent to a vertex $u_{2} \in V(H)$ with $c(u_{2})=j$. The diagram depicts the pair of low and high connector gadgets associated with color $j$ in the $u$-selection gadget that are connected to the $\{u_{1},u_{2}\}$-selection gadget. The remaining $(\ell-2)$ pairs of connector gadgets in the $u$-selection gadget are not depicted. 
  The rectangle boxes represent cliques and each ellipsis represents an independent set. 
  The dotted lines depict a complete set of edges between two sets of vertices.}
\label{fig:treewidth}
\end{center}
\end{figure}
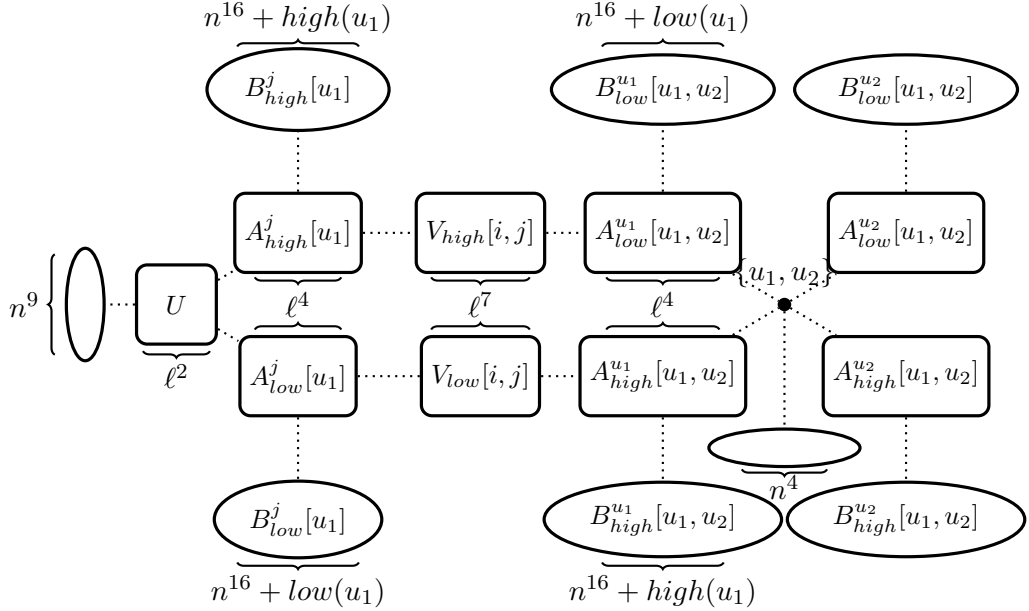

\noindent \textbf{Selection gadgets:} The graph $H$ consists of a selection gadget for each vertex and edge in $G$ (see Figure~\ref{fig:treewidth}): For a vertex $u \in V(G)$, we will construct a \emph{$u$-selection gadget} as follows: First we add a clique $U$ of size~$\A$ to $H$, and then we connect all the vertices of $U$ to an additional independent set of $\Y$ vertices, which we call the \emph{dummy vertices} of the $u$-selection gadget. We next connect $U$ to $(\ell-1)$ gadget pairs, one pair for each color $i \in \{1,\ldots,\ell\} \setminus \{c(u)\}$. Each pair consists of a low~order and a high~order connector gadget corresponding to $u$. We let $A^i_o[u]$ and $B^i_o[u]$ respectively denote the core and guard of the connector gadget \emph{associated} with color $i \in \{1,\ldots,\ell\} \setminus \{c(u)\}$ and of order $o \in \{low,high\}$. We connect $U$ to each connector gadget by adding all edges between all vertices of $U$ and $A^i_o[u]$, for each $i \in \{1,\ldots,\ell\} \setminus \{c(u)\}$ and $o \in \{low,high\}$.

For an edge $\{u_1,u_2\} \in E(G)$, we will construct a \emph{$\{u_1,u_2\}$-selection gadget} as follows: First we add a vertex which we denote by $\{u_1,u_2\}$ to $H$. We then connect $\{u_1,u_2\} \in V(H)$ to a low~order and a high~order connector gadget associated with $u_1$, and to a low~order and a high~order connector gadget associated with $u_2$, by adding all edges between vertex $\{u_1,u_2\} \in V(H)$ and the core vertices of these gadgets. We let $A^u_o[u_1,u_2]$ and $B^u_o[u_1,u_2]$ respectively denote the core and guard of the connector gadget corresponding to $u \in \{u_1,u_2\}$ of order $o \in \{low,high\}$ in the $\{u_1,u_2\}$-selection gadget. Finally, we connect $\{u_1,u_2\} \in V(H)$ to an additional set of $\X$ dummy neighbors of degree one in~$H$.\\ 

\noindent \textbf{Validation gadgets:} We next add the validation gadgets to $H$, one for each ordered pair of distinct colors $(i,j)$, $i \neq j$. For such a pair $(i,j)$, the $(i,j)$-validation gadget simply consists of two cliques $V_{low}[i,j]$ and $V_{high}[i,j]$, each of size $\C$. The validation is done through the connections of these two cliques to the remainder of the graph. Consider a $u$-selection gadget for a vertex $u \in V(G)$ of color $i$. We add all possible edges between $V_{low}[i,j]$ and $A^j_{low}[u]$, and all edges between $V_{high}[i,j]$ and $A^j_{high}[u]$. This is done for every vertex of color $i$. Consider next a $\{u_1,u_2\}$-selection gadget where $c(u_1)=i$ and $c(u_2)=j$. We add all possible edges between $V_{low}[i,j]$ and $A^{u_1}_{high}[u_1,u_2]$, and all possible edges between $V_{high}[i,j]$ and $A^{u_1}_{low}[u_1,u_2]$.  In this way, $V_{low}[i,j]$ is connected to low~order connector gadgets of vertex selection gadgets and to high~order connector gadgets of edge selection gadgets, and $V_{high}[i,j]$ is connected in the opposite way.\\

\noindent \textbf{CNC instance:} The graph $H$ of our CNC instance is thus composed of~$4\binom{\ell}{2}$ validation cliques which have~$\C$ vertices each,~$n$ vertex selection gadgets each of size $(\ell-1)(2\Z+2n+1+2\B)+\Y+\A$, and $m$ edge selection gadgets which have $2(2\Z+2n+1+2\B)+\X+1$ vertices each. We finish the description of our reduction by setting $k$, the size of the required critical node cut, to
$$
k:=\, \left(2(\ell-1)n+4m-8\binom{\ell}{2}\right)\cdot \B + \LA + \binom{\ell}{2},
$$
and setting $x$, the bound on the number of connected pairs, to
\begin{eqnarray*}
x & := & \left(n-\ell\right)(\Y+\A)(\Y+\A-1) +\left(m-\binom{\ell}{2}\right)(\X+1)\X +\\
& & 4\binom{\ell}{2}(2\Z+2n+1+\C+2\B)(2\Z+2n+\C+2\B).
\end{eqnarray*}

\begin{lemma}
The graph $H$ has treewidth at most $4\binom{\ell}{2}\C+\B+\A$.
\end{lemma}

\begin{proof}
In our proof we use two easy and well known facts about treewidth: The treewidth of a graph is the maximum treewidth of all its components, and adding $\alpha$ vertices to a graph of treewidth at most $\beta$ results in a graph of treewidth at most $\alpha + \beta$. Using these two facts we get that a connector gadget has treewidth at most $\B$, since we add $\B$ vertices to a graph of treewidth 0 (the independent set of vertices). From this we conclude that each selection gadget has treewidth at most $\B+\A$, since we either add a clique of size $\A$ or a single vertex to a graph whose connected components have treewidth bounded by~$\B$. Therefore, since~$H$ itself is constructed by adding $4\binom{\ell}{2} \cdot \C$ validation vertices to a graph whose connected components have treewidth at most $\B+\A$, the lemma follows. 
\end{proof}

\noindent \textbf{From a multicolored clique to a critical node cut:} Suppose $(G,c,\ell)$ has a solution, \emph{i.e.}, a multicolored clique $S$ of size $\ell$. Then one can verify that the cut $C \subseteq V(H)$ defined by
\begin{align*}
  C:= \{U : u \in S\} \cup \{\{u_1,u_2\} : u_1 \neq u_2 \in S\} & \cup \big\{v: v \in
  A^c_o[u], u \notin S\big\}\\ &  \cup \big\{v: v \in A^u_o[u_1,u_2], u_1
  \neq u_2 \notin S\big\}
\end{align*}
is of size $k$, and $H-C$ contains exactly three types of non-trivial connected components:
\begin{itemize}
\item $n-\ell$ components which include a clique $U$ of size $\A$ along with $\Y$ dummy vertices.
\item $m-\binom{\ell}{2}$ components which include a single vertex of $E(G)$ along with $\X$ dummy vertices.
\item $4\binom{\ell}{2}$ components which have $2\Z+2n+1+\C+2\B$ vertices each.
\end{itemize}
Thus, $H-C$ has exactly $x$ connected pairs, and $C$ is indeed a solution to $(H,k,x)$. \\

\noindent \textbf{From a critical node cut to a multicolored clique:} To complete the proof of Theorem~\ref{thm:w}, we show that if $(H,k,x)$ has a solution, \emph{i.e.}, a cut $C$ of size $k$ where $H-C$ has at most $x$ connected pairs, then $G$ has a multicolored clique of size $\ell$. We do this, using a few lemmas that restrict the structure of solutions to our \textsc{CNC} instance. The first one of these, Lemma~\ref{lem:w1} below, shows that we can restrict our attention to cuts which include only core vertices of connector gadgets and vertices of $V(G) \cup E(G)$.
\begin{lemma}
\label{lem:w1}%
If there is a solution to $(H,k,x)$, then there is a solution $C$ to this instance which includes no guard vertices, no dummy vertices, and no validation vertices of $H$.
\end{lemma}

\begin{proof}
Let $C$ be a solution to $(H,k,x)$. If $C$ includes any dummy vertex~$v$ of $H$, then since~$v$ is a vertex whose neighborhood is a clique, we can replace~$v$ with one of its neighbors (which is a non-dummy vertex) or if~$C$ contains all of the neighbors of~$v$, we remove~$v$ from~$C$. In both cases, the number of connected pairs in $H-C$ is not decreased. Similarly, if~$C$ includes guard vertices, these can be safely replaced with core vertices.

Next, we show that~$C$ cannot contain any validation clique completely. To this end, note that a core of a connector gadget which is not completely included in~$C$ contributes more than~$\ZZ$ connected pairs in~$H-C$. This can be seen by counting the number of connected pairs between a single core vertex and all of its guard neighbors. Thus, since $(16\binom{\ell}{2}+1)\ZZ > x$ assuming a sufficiently large $n$, the cut $C$ must include all but at most $16\binom{\ell}{2}$ cores of connector gadgets in $H$. But as each validation clique is of size $\C > 8\binom{\ell}{2}\B+\LA+\binom{\ell}{2}$ (for sufficiently large $\ell$), we have $k - \C < (2(\ell-1)n+4m-16\binom{\ell}{2})\B$, which means that if $C$ includes a validation clique it does not include enough cores. Thus, $C$ cannot completely contain any validation clique.


Finally, consider the case that $C$ contains a proper subset of some validation clique~$V_o[i,j]$ in $H$. Observe first that if the validation clique is not completely isolated in~$H-C$, then a vertex~$v\in C\cap V_o[i,j]$ can be safely replaced by a core vertex that is adjacent to~$V_o[i,j]$ as~$v$ is not a cut vertex in~$H-(C\setminus \{v\})$. Thus, the only remaining case is that all vertices that have a neighbor in~$V_o[i,j]$ are in~$C$. Then, deleting the vertices in~$V_o[i,j]$ removes at most~$\C(\C-1)$ connected pairs. By the choice of~$k$, and the number of core vertices,~$C$ cannot contain all core vertices. Consider a core vertex~$u\notin C$. Since~$C$ does not contain any guard vertices, adding~$u$ to~$C$ removes at least~$\Z>\C(\C-1)$ connected pairs. Thus, we can remove all vertices of~$V_o[i,j]\cap C$ from~$C$ and replace them by~$u$ without increasing the number of connected pairs in~$H-C$. Thus, there is a solution that contains no vertices of validation cliques. 
\end{proof}

Assume that  $(H,k,x)$ has a solution, and fix a solution $C$ as in Lemma~\ref{lem:w1}. By the definition of $k$, we know that the cut $C$ cannot include all connector gadgets. A connector gadget in $H-C$ induces a large number of connected pairs, at least $\ZZ$, due to the guard vertices of the gadget. Let us therefore call a connected component in $H-C$ \emph{huge} if it contains at least $\ZZ$ connected pairs. The next lemma shows that there can only be a certain number of these huge components in $H-C$, and reveals some restriction on any solution cut $C$. We call a maximal non-empty (but not necessarily proper) subset of a core in $H-C$ a \emph{partial core}.

\begin{lemma}
\label{lem:w2}%
If $C$ is a solution to $(H,k,x)$ as in Lemma~\ref{lem:w1}, then $C$ includes $(2(\ell-1)n+4m-8\binom{\ell}{2})$ cores. Furthermore, there are precisely $4\binom{\ell}{2}$ huge components in $H-C$, each of which consists of a validation clique, two partial cores, and the two guard sets of the partial cores.
\end{lemma}

\begin{proof}
Let $A_1,\ldots,A_t$ denote all partial cores in $H-C$. Note that since each core is of size $\B > \LA + \binom{\ell}{2}$ (for sufficiently large $\ell$), the cut $C$ can include at most $(2(\ell-1)n+4m-8\binom{\ell}{2})$ complete cores by definition of~$k$, and so $t \geq 8\binom{\ell}{2}$. By Lemma~\ref{lem:w1}, the graph $H-C$ contains all $4\binom{\ell}{2}$ validation cliques. Let $Q_1,\ldots,Q_s$ denote the components in $H-C$ that contain at least one validation clique, and let $q_i:=|Q_i|-1$ for each $i$, $1 \leq i \leq s$. Observe that for any huge component $Q$ in $H-C$, we have $Q \in \{Q_1,\ldots,Q_s\}$.

Now, since the total number of validation cliques is $4\binom{\ell}{2}$, we have $s \leq 4\binom{\ell}{2}$, and the total number of connected pairs in all the $Q_i$'s is lower bounded by $\sum_{i=1}^s q_i^2$. Note that each partial core $A_j$ belongs to some $Q_i$ and contributes at least $\Z+1$ vertices to its size (accounting for a single vertex of $A_j$ and all its guard neighbors), and therefore at least $\ZZ$ connected pairs. It can now be seen that since $\sum_{i=1}^{s} q_i^2$ is concave and symmetric then it is minimized when the number of addends is as large as possible and all of the addends are of equal size. This happens when $s=4\binom{\ell}{2}$ and each $Q_i$ includes exactly two~$A_j$'s, giving us $\sum_{i=1}^{s} q_i^2=\sum_{i=1}^{s} ((2\Z)^2)+o(\ZZ)=16\binom{\ell}{2}\ZZ+o(\ZZ)$. If $s<4\binom{\ell}{2}$ or there is one $Q_i$ that contains more then two~$A_j$'s, then the sum will be at least $(16\binom{\ell}{2}+1)\ZZ > x$. It follows that there are exactly $4\binom{\ell}{2}$ huge components, that each have two~$A_j$'s. These huge components contribute altogether at least $16\binom{\ell}{2}\ZZ$ connected pairs.

We have thus established that there are $4\binom{\ell}{2}$ huge
components in $H-C$, and each includes a validation clique, two
partial cores, and the guard sets adjacent to these partial cores
which are not in $C$ according to Lemma~\ref{lem:w1}. To see that the
huge components contain nothing else recall first that the
overall number of connected pairs in these huge components is at
least~$16\binom{\ell}{2}\ZZ$. Thus, the number of further additional
connected pairs in~$H-C$ is at most~$x - 16\binom{\ell}{2}\ZZ = n\cdot \YY+ o(\YY) < 2n^{19}$.  Now, if $A$ contains other
vertices, then by construction it must contain either a vertex from a
clique~$U$ corresponding to a vertex~$u$ of~$G$, or a
vertex~$\{u,u'\}$ corresponding to an edge of~$G$. In either of these
cases, this additional vertex is adjacent to at least~$\X$ dummy
vertices, implying that~$Q$ has an additional number of~$\X\cdot \Z=n^{20} > 2n^{19}$
connected pairs, a contradiction. 
\end{proof}

Slightly smaller than huge components are \emph{large components} in $H-C$ which have at least $\YY$ connected pairs and fewer than $\ZZ$ connected pairs. Further smaller are \emph{big components} which have at least $\XX$ connected pairs, and less than $\YY$ connected pairs.

\begin{lemma}
\label{lem:w3}%
If $C$ is a solution to $(H,k,x)$ as in Lemma~\ref{lem:w1}, then $C$ includes exactly $\ell$ cliques $U_1,\ldots,U_\ell$ corresponding to vertices $u_1,\ldots,u_\ell \in V(G)$, and there are precisely $n-\ell$ large components in $H-C$.
\end{lemma}

\begin{proof}
Note that $x=16\binom{\ell}{2}\ZZ + (n-\ell)\YY + o(\YY)$. By Lemma~\ref{lem:w2}, we know that $H-C$ contains $4\binom{\ell}{2}$ huge components, and so these already account for $16\binom{\ell}{2}\ZZ$ connected pairs in $H-C$. For every $u \in V(G)$, if the clique corresponding to $U$ is not completely contained in $C$, then there is a large component corresponding to $u$ in $H-C$, since by Lemma~\ref{lem:w1}, all $\Y$ dummy neighbors of $U$ are existent in $H-C$. Furthermore, any large component in $H-C$ is of this form. Thus, if $C$ contains $\ell' < \ell$ cliques corresponding to vertices of $G$, then the number of connected pairs in $H-C$ is at least $16\binom{\ell}{2}\ZZ + (n-\ell')\YY > 16\binom{\ell}{2}\ZZ + (n-\ell)\YY + o(\YY) = x$, a contradiction. Moreover, by our choice of~$k$, the cut $C$ cannot include $(2(\ell-1)n+4m-8\binom{\ell}{2})$ cores (as is necessary by Lemma~\ref{lem:w2}) and more than $\ell$ such cliques $U$, since $(2(\ell-1)n+4m-8\binom{\ell}{2})\B+(\ell+1)\A > k$. 

\end{proof}

\begin{lemma}
\label{lem:w4}%
If $C$ is a solution to $(H,k,x)$ as in Lemma~\ref{lem:w1}, then $C$ includes exactly $\binom{\ell}{2}$ vertices which correspond to edges in $G$, and there are precisely $m-\binom{\ell}{2}$ big components in $H-C$.
\end{lemma}

\begin{proof}
Let us call each element in the set $\{U \subset V(H) : u \in V(G)\} \cup \{\{u,u'\} \in V(H) : \{u,u'\} \in E(G)\}$ a \emph{$G$-element}. Thus, each $G$-element belongs to its unique selection gadget in $H$, and corresponds to either a vertex or an edge of $G$. Moreover, each core is adjacent to exactly one~$G$-element. By Lemma~\ref{lem:w3} we know that $C$ contains $\ell$ $G$-elements corresponding to vertices of $G$. We next argue that it also contains $\binom{\ell}{2}$ $G$-elements corresponding to edges of~$G$.

Consider a huge component~$Q$ in $H-C$. By Lemma~\ref{lem:w2},~$Q$ contains two partial cores $A$ and $A'$ and~$Q$ does \emph{not} contain the unique~$G$-element that is adjacent to the two partial cores. Thus, the~$G$-element neighbors of exactly~$8\binom{\ell}{2}$ partial cores are contained in~$C$. The set of cliques~$U_1,\ldots,U_\ell \subseteq C$ promised by Lemma~\ref{lem:w3} accounts for at most $2(\ell-1)\cdot\ell=4\binom{\ell}{2}$ such cores, as each~$U_i$ has exactly~$2(\ell-1)$ neighboring cores in~$H$. Notice that by the choice of~$k$, after accounting for the vertices in $C$ required by Lemma~\ref{lem:w2} and Lemma~\ref{lem:w3}, the remaining number of vertices is~$\binom{\ell}{2}$. Another $G$-element representing a vertex requires $\A>\binom{\ell}{2}$, thus all remaining deleted~$G$-elements correspond to edges of~$G$. Now observe that each of them can account for at most four partial cores as they have exactly four neighboring cores in~$H$. Consequently, the number of deleted~$G$-elements that correspond to edges in~$G$ is at least~$\binom{\ell}{2}$. By the choice of~$k$, it is thus exactly~$\binom{\ell}{2}$.
\end{proof}

\begin{lemma}
The set of vertices $u_1,\ldots,u_\ell$ specified in Lemma~\ref{lem:w3} induces a multicolored clique in $G$.
\end{lemma}

\begin{proof}
Lemma~\ref{lem:w2}, Lemma~\ref{lem:w3}, and Lemma~\ref{lem:w4} together state that $C$ includes at least $(2(\ell-1)n+4m-8\binom{\ell}{2}) \cdot \B$ core vertices, at least $\LA$ vertices in cliques corresponding to vertices of $G$, and at least $\binom{\ell}{2}$ vertices corresponding to edges of $G$. By our selection of $k$, all these lower bounds are in fact equalities. Thus, all but $\ell$ cliques $U$, $u \in V(G)$, are present in $H-C$, and all but $\binom{\ell}{2}$ edges of $G$ are present in $H-C$. All these vertices contribute at least $\left(n-\ell\right)(\Y+\A)(\Y+\A-1) +\left(m-\binom{\ell}{2}\right)(\X+1)\X$ connected pairs in $H-C$, due to their dummy neighbors. Thus, by definition of $x$, the total number of connected pairs from huge components in $H-C$ is $4\binom{\ell}{2}(2\Z+2n+1+\C+2\B)(2\Z+2n+\C+2\B)$.

Now, note that according to Lemma~\ref{lem:w2}, $H-C$ includes exactly $8\binom{\ell}{2}$ partial cores with no neighboring $G$-elements. The set of cliques~$U_1,\ldots,U_\ell \subseteq C$ ,promised by Lemma~\ref{lem:w3}, accounts for at most $2(\ell-1)\cdot\ell=4\binom{\ell}{2}$ partial cores, only if each vertex is of a different color. Moreover, the $\binom{\ell}{2}$ deleted~$G$-elements that correspond to edges in~$G$, promised by Lemma~\ref{lem:w4}, accounts for at most $4\binom{\ell}{2}$ partial cores, only if each edge corresponds to a different pair of colors. Consequently, the only way to remove the required number of neighboring $G$-elements is if these upper bounds are met with equality. Thus, the vertices and edges corresponding to the removed $G$-elements are of different colors, as required in a multicolored clique.


Finally, observe that due to the fact that we have accounted for all the vertices in $C$, it is clear that each huge component consists of two complete (\emph{i.e.} non-partial) cores. Thus, the size of each of these huge components is $2\Z+\C+2\B+high(u_{1})+low(u_{1}')$ for $u_{1},u_{1}'\in V(G)$. Therefore, the only way for the total number of connected pairs in all huge components to not exceed $4\binom{\ell}{2}(2\Z+2n+1+\C+2\B)(2\Z+2n+\C+2\B)$ is if all huge components have equal size, \emph{i.e.}, exactly $(2\Z+2n+1+\C+2\B)$~vertices each. But this can happen only if we have $u_{1}=u_{1}'$ in the pair of connector guards~$B^i_o[u_{1}]$ and~$B^{u_{1}'}_{\bar{o}}[u_{1}',u_2]$, in each huge component of~$H-C$, as this is the only way for the guard vertices to sum up to~$2\Z+2n+1$. Consequently, the set of $\binom{\ell}{2}$ edges selected in $G$ are edges between $u_1,\ldots,u_\ell$ implying that they indeed form a clique. 
\end{proof}

\section{Parameter \boldmath{$w+x$}}
\label{sec:w+x}
If we combine the treewidth parameter~$w$ with the parameter for the
number of connected pairs~$x$, then we obtain fixed-parameter
tractability. This can be derived via an optimization variant of Courcelle's theorem due to~\cite{arnborg1991easy}. 
Using tree decompositions, we obtain a more efficient algorithm.

A \emph{nice tree decomposition}~\cite{36} of a graph $G$ is a pair $\langle \mathcal{X},\mathcal{T} \rangle$, where each element $X \in \mathcal{X}$ (called a \emph{bag}) is a subset of $V(G)$, and $\mathcal{T}$ is a rooted tree over $\mathcal{X}$. The pair $\langle \mathcal{X},\mathcal{T} \rangle$ is required to satisfy the following conditions:
\begin{enumerate}
\item $\bigcup_{X \in \mathcal{X}}  X=V(G)$.
\item For every edge $\{u,v\}\in E(G)$, there is an $X \in \mathcal{X}$ with $\{u,v\}\subseteq X$.
\item For all $X,Y,Z \in \mathcal{X}$, if $Y$ lies on the path between $X$ and $Z$ in $\mathcal{T}$, then $X \cap Z\subseteq Y$.
\item There are $4$ types of bags:
    \begin{enumerate}
    \item A \emph{leaf bag} $X$ which has no children in $\mathcal{T}$ and contains a single vertex $v \in V(G)$.
    \item An \emph{introduce bag} $X$ which has a single child $Y$ in $\mathcal{T}$ with $X = Y \cup \{v\}$ for some vertex $v \notin Y$.
    \item A \emph{forget bag} $X$ which has a single child $Y$ in $\mathcal{T}$ with $X = Y \setminus \{v\}$ for some vertex $v \in Y$.
    \item A \emph{join bag} $X$ which has two children $Y$ and $Z$ in $\mathcal{T}$ with $X=Y=Z$.
    \end{enumerate}
\end{enumerate}
Note that Conditions~$1-3$ define a tree decomposition and their combination with Condition~$4$ defines a nice tree decomposition. The width of a tree decomposition is the number of elements in the largest bag minus $1$. The treewidth $w$ of $G$ is the minimum width of all the possible tree decompositions of $G$. For a given graph $G$ with treewidth $w$, one can obtain its nice tree decomposition in $f(w) \cdot n^{O(1)}$ with $O(wn)$ bags~\cite{36,bodlaender93linear}. Thus, in proving the main result of this section, stated in the theorem below, we can assume that we are given as input a nice tree decomposition $\langle \mathcal{X},\mathcal{T} \rangle$ of width $w$ (and $O(wn)$ bags) of our input graph $G$.

\begin{theorem}
\label{thm:w+x}%
The \textsc{Critical Node Cut} problem is \textnormal{FPT} with respect to $w+x$.
\end{theorem}

Let $X$ be a bag from our nice tree decomposition, and let $G_{X}$ denote the subgraph of $G$ induced by the bag $X$ and all of its descendants in $\mathcal{T}$. We build a table for $X$. Each entry in this table is denoted by $T_{X}[k',x',X_{0},X_{1},\ldots ,X_{t},n_{1},\ldots ,n_{t}]$, where $k' \leq k$, $x' \leq x$, $n_i \leq x$ and $X_i \subseteq X$ for all $i$. The entry can either equal $0$ or $1$. It equals $1$ iff there exist $k'$ vertices $C$ in $G_{X}$, $X_{0}\subseteq C$, such that $G_{X} - C$ has at most $x'$ connected pairs and is separated into components $R_{1},R_{2},\ldots ,R_{t}$, with $X\cap R_{i}=X_{i}$ and $|R_{i}|=n_{i}$ for each $i$, $1 \leq i \leq t$. If there is no such solution then the entry equals $0$. Thus, an entry with value 1 corresponds to a partial solution that splits the bag $X$ in a very particular way. Note that $R_{1},R_{2},\ldots ,R_{t}$ are only the components that intersect the bag $X$. Our algorithm calculates the tables of each node in the decomposition in a bottom-up fashion. Clearly, if each entry is computed correctly then one can infer whether there exists a solution to the \textsc{CNC} instance by examining the table at the root. Note that for each bag the size of the table is $O(nx(w+x)^{w})$, thus if we show that calculating an entry can be done in FPT time, then we prove that our algorithm altogether runs in FPT time.

We will next show how to calculate $T_{X}$ for each possible type of bag $X$. If $X$ is a leaf bag, then $X$ is composed of one vertex only, and therefore the computation in this case is trivial. If $X$ is a forget bag with a child $Y$ such that $Y=X\cup \{v\}$ for $v \notin X$, then the entry $T_{X}[k',x',X_{0},X_{1},\ldots ,X_{i},\ldots ,X_{t},n_{1},\ldots ,n_{t}]$ will equal $1$ if and only if there exists in the table of $Y$ an entry $T_{Y}[k',x',X_{0},X_{1},\ldots ,X_{i}\cup \{v\},\ldots ,X_{t},n_{1},\ldots ,n_{t}]$ that equals $1$. This is correct because the only difference between the entries is that $v\notin X$, therefore it is excluded from its partition, yet it is still counted by the value $n_{i}$ as a member of the corresponding component. To complete the proof of Theorem~\ref{thm:w+x}, we show in the next two lemmas that we can efficiently calculate each entry in $T_X$ also if $X$ is an introduce or a join bag.

\begin{lemma}
\label{lem:introduce}
If $X$ is an introduce bag with child $Y$, then, given the table $T_Y$, an entry in $T_X$ can be calculated in~$|T_y|^{O(1)}$ time.
\end{lemma}

\begin{proof}
Let $v$ be the single vertex in $X \setminus Y$, and consider an arbitrary entry $$T_X[k',x',X_{0},X_{1},\ldots ,X_{t},n_{1},\ldots ,n_{t}]$$ in the table of $X$. There are three possible cases:
\begin{enumerate}
\item $v\in X_{0}$. The entry in $T_{X}$ will equal $1$ iff in $T_Y$ we have
    $$
    T_{Y}[k'-1,x',X_{0}\setminus\{v\},X_{1},\ldots ,X_{t},n_{1},\ldots ,n_{t}] = 1.
    $$
\item $v \notin X_{0}$ and $v$ is not adjacent to any vertex in $G_{Y}$. The only type of entry $T_{X}$ that might have value $1$ is an entry where $X_{i} = \{v\}$ for some $i\geq 1$, and $n_{i}=1$. If this is the case then the current entry in $T_{X}$ will be $1$ iff we have
    $$
    T_{Y}[k',x',X_{0},\ldots ,X_{i-1},X_{i+1},\ldots ,X_{t},n_{1},\ldots ,n_{i-1},n_{i+1},\ldots ,n_{t}]=1.
    $$
\item $v\notin X_{0}$ and $v$ is adjacent to vertices of $Y$. By the properties of a tree decomposition, in $G_X$ the vertex $v$ can only have neighbors from $Y$. Let $y_{1},\ldots ,y_{a} \in Y$ denote these neighbors. Then the only entries in $T_X$ that might have value 1 are entries with a subset $X_i \subseteq X$ that includes $v$ and all his neighbors. If the current entry in~$T_{X}$ is as such, then it will have value $1$ iff
    $$
    T_{Y}[k',x'',X_{0},X_{1},\ldots ,X_{i-1},Y_1,\ldots,Y_b,X_{i+1},\ldots ,X_{t},n_{1},\ldots ,n_{i-1},m_1,\ldots,m_b,n_{i+1},\ldots ,n_{t},]=1,
    $$
    for an entry in $T_Y$ where $Y_1,\ldots,Y_b$ are precisely the subsets of $Y$ that include neighbors of~$v$, $X_i =\{v\} \cup \bigcup_j Y_j$, $x''=x'- 2\sum_j m_j - \sum_j m_j \sum_{k\neq j} m_k$, and~$n_i=1+\sum_j m_j$.
\end{enumerate}

The correctness of the first two cases is easy to see. The reason that Case~$3$ is correct is that if the entry from $Y$ exists then adding $v$ will connect the components $R_1,\ldots,R_b$ in $G_Y$ corresponding to $Y_1,,\ldots ,Y_b$ above into one component $X_{i}$. The size of the new component will be the sum of the sizes of the previous components plus $1$ (because of $v$). The number of connected pairs added in the process is the number of connected pairs between $v$ and all vertices in $\bigcup_j R_j$ ($2\sum_{j=1}^{b} m_j$), and the number of connections between vertices in different components ($\sum_{j=1}^{b} m_j \sum_{k\neq j} m_k$). Conversely, any given entry of value 1 in $T_X$ must correspond to an entry from~$T_Y$ as stated above. Since the computation of the entry in $T_X$ is clearly polynomial in the size of $T_Y$, the lemma follows. 
\end{proof}

\begin{lemma}
If $X$ is a join node with children $Y$ and $Z$, then, given the tables $T_Y$ and $T_Z$, an entry in $T_X$ can be calculated in~$(|T_y|+|T_z|)^{O(1)}$ time.
\label{lem:join}
\end{lemma}

\begin{proof}
Recall that by definition we have $X=Y=Z$. Let $T_X[k',x',X_{0},X_{1},\ldots ,X_{t},n_{1},\ldots ,n_{t}]$ be an arbitrary entry in the table of $X$. This entry equals $1$ iff there exist in $T_Y$ and $T_Z$ the entries $T_{Y}[k'_Y,x'_Y, Y_{0},Y_{1},\ldots ,Y_{p},n'_1,\ldots ,n'_{p}]$ and $T_{Z}[k'_Z,x'_Z,Z_{0},Z_{1},\ldots ,Z_{q},n''_1,\ldots ,n''_{q}]$ that equal $1$ and satisfy the following conditions:
\begin{enumerate}
\item $X_{0}=Y_{0}\cup Z_{0}$ and also $k=k'_Y+k'_Z- |Y_{0}\cap Z_{0}|$.
\item We define a relation $\approx$ on the vertices of $X$ given by $u \approx v$  iff $\{u,v\}\in Y_{i}$ or $\{u,v\}\in Z_{j}$ for $i\in \{1,\ldots,p\}$ and $j \in \{1,\ldots,q\}$. The partition $X_{1},\ldots ,X_{t}$ is required to be defined by the equivalence classes of the transitive closure of this relation.
\item The size $n_i$ of each component corresponding to $X_i$ match up according to the differences between the new partition and the old ones using the exclusion-inclusion principle. That is, if $X_i = Y_{j_1} \cup \ldots \cup Y_{j_\alpha} \cup Z_{k_1} \cup \ldots \cup Z_{k_\beta}$, then let $A_s = Y_{j_s}$ for $s \in \{1,\ldots,\alpha\}$ and $A_s = Z_{k_{s-\alpha}}$ for $s \in \{\alpha+1,\ldots,\alpha+\beta\}$, and we require that
    $$
    n_i= \sum_{\ell=1}^{\alpha} n'_{j_\ell} + \sum_{\ell=1}^{\beta} n''_{k_\ell} + \sum_{\substack{L \subseteq \{1,\ldots,\alpha+\beta\},\\ |L| \geq 2}} (-1)^{|L|-1}\left| \bigcap_{\ell \in L} A_\ell \right|.
    $$
    \item The bound on the number of connected pairs $x'$ adds up in the correct way. That is,
    $$
    x'=x'_Y+x'_Z + \sum_{i=1}^t n_i(n_i-1) - \sum_{j=1}^p n'_{j}(n'_j-1) - \sum_{k=1}^q n''_{k}(n''_k-1).
    $$
\end{enumerate}

Note that for the given entry in $T_X$, we can verify whether there exist appropriate entries in $T_Y$ and $T_Z$ that satisfy the four requirement above in polynomial time with respect to the sizes of this tables. Furthermore, it can be readily verified that if $R^Y_1\ldots,R^Y_p$ are connected components in $G_Y$ corresponding to a partition $\{Y_0,Y_1,\ldots,Y_p\}$ of $Y$, and $R^Z_1\ldots,R^Z_q$ are connected components in $G_Z$ corresponding to a partition $\{Z_0,Z_1,\ldots,Z_q\}$ of $Z$, then in $G_X$ we will have $t$ connected components $R^X_1\ldots,R^X_t$ that correspond to the partition $\{X_0,X_1,\ldots,X_t\}$ defined above. This follows directly from the fact that connectivity is an equivalence relation, and the fact that a component $R^Y_j$ can intersect a component $R^Z_k$ only at vertices of $X$. Thus, since all other requirements are direct corollaries of this fact, the existence of two such entries in $T_Y$ and $T_Z$ imply that the current entry in $T_X$ equals 1. The converse implication, that is, the fact that if the current entry of $T_X$ equals 1 there must exist two entries in $T_Y$ and $T_Z$ which equal 1 and satisfy the above requirements, follows along the same lines. 
\end{proof} 

\section{Parameter \boldmath{$y$}}
\label{sec:y}
%

Finally, we consider the \textsc{CNC} problem parameterized by $y$. We will show that the problem is FPT under this parameterization but has no polynomial kernel even for the aggregate parameterization of $k+y+w$; the proofs are deferred to the appendix.
\begin{theorem}
The \textsc{Critical Node Cut} problem is FPT with respect to $y$.
\label{thm:y}
\end{theorem}

\begin{proof}
Let $(G,k,y)$ be an instance of \textsc{CNC}. Observe that if one of the components was larger than $y$, then removing one vertex from this component already causes the removal of at least $y$ connected pairs. Moreover, if $k$ was larger than $y$ (in fact, $y/2$) then removing any $k$ arbitrary non-isolated vertices has the same effect. Thus, the interesting case occurs when $k < y$ and each component of $G$ has size at most $y$, and we assume henceforth throughout the proof that this is in fact the case.

Our algorithm proceeds as follows, herein let~$G_1,\ldots, G_t$ denote the connected components of~$G$:
\begin{enumerate}
\item For each component~$G_i$ of~$G$ and each~$k'$,~$1\le k'\le k$,
  compute by brute-force the maximum number of connected pairs
  in~$G_i$ that can be removed by deleting exactly~$k'$ vertices
  in~$G_i$. Let~$T[i,k']$ denote this number.
\item For increasing~$i$, compute the maximum number of connected
  pairs that can be removed by deleting exactly~$k'$ vertices in the
  components~$G_1,\ldots, G_i$. Let~$Q[i,k']$ denote this number. For~$i=1$,
  we have~$Q[1,k']=T[1,k']$. For~$i>1$, we have $$Q[i,k']=
  \max_{k''\le k'} Q[i-1,k'']+T[i,k'-k''].$$
\item If~$Q[t,k]<y$ return NO; otherwise, return YES.
\end{enumerate}
Correctness of the algorithm is rather obvious: optimal solutions for
different components can be combined since the connected pairs are
only contained within each component. The running time
is~$O(2^{y}\cdot y^2\cdot n)$: Each component has at most~$y$ vertices,
thus there are~$O(2^y)$ possibilities to consider in the brute-force
step. For each possibility, computing the size of the remaining
connected components can be done in~$O(y^2)$ time. The dynamic
programming in the second step of the algorithm is then performed
for~$t\le n$ different values of~$i$. For each value of~$i$,~$k^2\le
y^2$ possible combinations of~$k'$ and~$k''$ are considered.
\end{proof}
Using the cross-composition technique of Bodlaender et
al.~\cite{BJK14}, we now show that parameter $y$ does not seem to be
useful when considering polynomial kernelization for \textsc{CNC}. We
give the definition directly applied to \textsc{CNC}.
\begin{definition}[\cite{BJK14}]
\label{Definition: Cross-Composition}%
A \emph{cross-composition algorithm} for \textsc{CNC} parameterized by
$k+y+w$ is a polynomial time algorithm that receives as input a
sequence of instances $I_1,I_2, \ldots, I_t$ of a problem~$L$ which
are equivalent under a polynomial equivalence relation and outputs an
instance $(G,k',y')$ of \textsc{CNC} such that:
\begin{itemize}
\item $(G, k',y')$ is a yes-instance of \textsc{CNC} iff some~$I_i$ is
  a yes-instance of~$L$.
\item $k' + y' + w \le \max_{i=1}^t |I_i|^{O(1)}+\log t$, where $w$ is
  the treewidth of $G$.
\end{itemize}
\end{definition}
The problem~$L$ will be \textsc{Clique} and the polynomial equivalence
relation will be that all instances have the same number of vertices
and edges. As shown by Bodlaender et al.~\cite{BJK14}, if~$L$ is
NP-hard, then the existence of a cross-composition to~\textsc{CNC}
implies the following theorem.
\begin{theorem}
The \textsc{Critical Node Cut} problem parameterized by $k+y+w$ has no polynomial kernel unless the polynomial hierarchy collapses.
\label{thm:y+k+w}
\end{theorem}
\begin{proof}
  We present a cross-composition from
  \textsc{Clique}. Let~$(G_1,\ell), ,...,(G_{t},\ell)$ be a set of~$t$
  \textsc{Clique} instances each with~$n$ vertices and~$m$
  edges. Assume without loss of generality that~$\ell>3$, that~$n > k^4$ and that~$m\ge n$. Now, first
  transform each~$G_i$ into an equivalent \textsc{CNC} instance
  exactly as in the proof of Theorem~\ref{thm:k}. That is, replace
  each edge in~$G_i$ by~$n$ parallel edges and then subdivide these
  edges. Call the new vertices dummy vertices. Then, make all vertices
  that are nonadjacent in~$G_i$ a clique and call the resulting
  graph~$H_i$ and let~$H$ be the disjoint union of
  all~$H_i$'s. Finally, let~$k:=\ell$ and~$$y:=k(k-1) +
  2k(N-k)+\binom{k}{2}n\cdot \left(\binom{k}{2}n-1\right) +
  2\binom{k}{2}n\left(N-k-\binom{k}{2}n\right)$$ where~$N:=|V(H_1)|$
  (by construction all~$H_i$'s have the same number of
  vertices). Clearly, the parameters~$k$ and~$y$ and~$N$ are bounded
  by polynomial function in~$n$. Moreover, since the resulting graph
  is a disjoint union of graphs with~$N$ vertices, its treewidth~$w$
  is also bounded by a polynomial function in~$n$. It remains to show
  that one of the $G_i$'s has a clique of size $\ell$ if and only if
  the graph $H$ has $k$ vertices whose removal deletes $y$ connected
  pairs in $H$.

  First, suppose $C$ is a clique of size $\ell$ in some~$G_i$.  and
  let~$D(C)$ denote the~$\binom{k}{2}n$ many dummy vertices that have
  both neighbors in~$C$. Removing $C$ in $H$ results in the deletion
  of $y$ connected pairs from $H$: A total of
  \begin{itemize}
  \item $k(k-1)$ connected pairs which involve only vertices of $C$,
  \item $2k(N-k)$ pairs which involve one vertex from~$C$ and one
    vertex from~$V(H_i)\setminus C$,
  \item $\binom{k}{2}n\cdot \left(\binom{k}{2}n-1\right)$ which
    involve only vertices from~$D(C)$, and
  \item $2\binom{k}{2}n\left(N-k-\binom{k}{2}n\right)$ connected pairs
    that involve one vertex of~$D(C)$ and one vertex
    of~$|V(H_i)|\setminus (C\cup D(C))$.
  \end{itemize}

  Conversely, suppose that $C$ is a cut that removes $y$ connected
  pairs in $H$. If $C$ contains a subset $C'\subseteq C$ of dummy
  vertices, then we can replace $C'$ with an arbitrary equally sized
  set of non-dummy vertices without decreasing $y$. Thus, we can
  assume that $C$ contains only non-dummy vertices. Furthermore,
  notice that when we remove a non-dummy vertex~$v$ (\emph{i.e.}, a
  vertex of some~$G_i$), then the only connected pairs that are
  deleted are the ones which either involve~$v$ or possibly dummy
  vertices that are neighbors of~$v$. This is because every pair of
  vertices from each $G_i$ is either connected by an edge or by an
  edge-gadget in $H_i$. Thus, the number of deleted connected pairs
  by~$k$ vertex deletions is at most $2k\cdot (N-1) + 2qn\cdot (N-3)$,
  where~$q$ is the number of times that we isolate~$n$ dummy vertices
  by deleting their two neighbors in~$H$. Now consider a set of~$k$
  vertices such that at least two vertices are from two
  different~$H_i$'s. The number~$q$ of times that we isolate~$n$ dummy
  vertices by deleting their two neighbors in~$H$ is at
  most~$\binom{k-1}{2}$. Thus, the number of deleted connected pairs
  is at most
  \begin{align*}
    y'= 2k(N-1) +2\binom{k-1}{2}n(N-3) & =2k(N-1)+2\binom{k-1}{2}n N
    - 6\binom{k-1}{2}\\ 
    & < 2kN + 2\binom{k}{2}n N
    -2knN.
  \end{align*}
  Moreover, observe that~$$y > 2kN + 2\binom{k}{2}nN - k^2 -k -
  2\binom{k}{2}n\left(k+\binom{k}{2}n\right).$$ Now for sufficiently
  large~$k$,~$y'$ is smaller than~$y$ since~$N> n\cdot m \ge n^2>k^4
  n$. Therefore, any solution deletes vertices from exactly
  one~$H_i$. Consequently, the number of deleted pairs that involve
  the removed vertices is exactly~$k(k-1)+2k\cdot (N-k)$ since the~$k$
  removed vertices are all from the same connected component
  of~$H$. Now the only way to delete $\binom{k}{2}n\cdot
  \left(\binom{k}{2}n-1\right) +
  2\binom{k}{2}n\left(N-k-\binom{k}{2}n\right)$ further connected
  pairs in $H$ is to remove $k$ vertices from~$H_i$ which are pairwise
  connected by edge-gadgets. By construction, these vertices
  correspond to $\ell=k$ vertices that form a clique in $G_i$. 
\end{proof}

\section{Discussion}
\label{sec:discussion}

We considered a natural graph cut problem called \textsc{Critical Node Cut (CNC)} under the framework of parameterized complexity. 
The only parameterization left open in our analysis is the parameter $w+k$, and so the first natural question left open in the paper is whether \textsc{CNC} is fixed-parameter tractable under this parameterization (we know it is unlikely that it admits a polynomial kernel). Other natural parameters could also be considered. For example, it would be interesting to see how parameters \emph{maximum degree} and \emph{pathwidth} affect the parameterized complexity of \textsc{CNC}. Finally, one can consider the edge variant of the problem (where one is required to delete edges instead of vertices) and the directed variant of the problem. Many of our results do not hold for these two variants.

\subparagraph{Acknowledgments.} 
The research leading to these results has received funding from the People Programme (Marie Curie Actions) of the European Union's Seventh Framework Programme (FP7/2007-2013) under REA grant agreement number 631163.11, and by the ISRAEL SCIENCE FOUNDATION (grant No. 551145/).

\bibliographystyle{plain}
\bibliography{biblo}

\end{document}